\documentclass[aps,pra,reprint,superscriptaddress,showpacs]{revtex4-1}

\usepackage{graphicx}
\usepackage{amsmath}
\usepackage{float}
\usepackage{amssymb}

\begin{document}


\title{Weak measurement of elliptical dipole moments by C point splitting}



\author{Sergey Nechayev}\thanks{These two authors contributed equally}
\affiliation{Max Planck Institute for the Science of Light, Staudtstr. 2, D-91058 Erlangen, Germany}
\affiliation{Institute of Optics, Information and Photonics, University Erlangen-Nuremberg, Staudtstr. 7/B2, D-91058 Erlangen, Germany}
\author{Martin Neugebauer}\thanks{These two authors contributed equally}
\affiliation{Max Planck Institute for the Science of Light, Staudtstr. 2, D-91058 Erlangen, Germany}
\affiliation{Institute of Optics, Information and Photonics, University Erlangen-Nuremberg, Staudtstr. 7/B2, D-91058 Erlangen, Germany}
\author{Martin Vorndran}
\affiliation{Max Planck Institute for the Science of Light, Staudtstr. 2, D-91058 Erlangen, Germany}
\affiliation{Institute of Optics, Information and Photonics, University Erlangen-Nuremberg, Staudtstr. 7/B2, D-91058 Erlangen, Germany}
\author{Gerd Leuchs}
\affiliation{Max Planck Institute for the Science of Light, Staudtstr. 2, D-91058 Erlangen, Germany}
\affiliation{Institute of Optics, Information and Photonics, University Erlangen-Nuremberg, Staudtstr. 7/B2, D-91058 Erlangen, Germany}
\author{Peter Banzer}
\email[]{peter.banzer@mpl.mpg.de}
\homepage[]{http://www.mpl.mpg.de/}
\affiliation{Max Planck Institute for the Science of Light, Staudtstr. 2, D-91058 Erlangen, Germany}
\affiliation{Institute of Optics, Information and Photonics, University Erlangen-Nuremberg, Staudtstr. 7/B2, D-91058 Erlangen, Germany}



\date{\today}

\begin{abstract}
We investigate points of circular polarization in the far field of elliptically polarized dipoles and establish a relation between the angular position and helicity of these C points and the dipole moment. In the case of highly eccentric dipoles, the C points of opposite handedness exhibit only a small angular separation and occur in the low intensity region of the emission pattern. In this regard, we introduce an optical weak measurement approach that utilizes the transverse electric (azimuthal) and transverse magnetic (radial) far-field polarization basis. Projecting the far field onto a spatially varying post-selected polarization state reveals the angular separation and the helicity of the C points. We demonstrate the applicability of this approach and determine the elliptical dipole moment of a particle sitting on an interface by measuring the C points in its far field. 
\end{abstract}


\maketitle

\enlargethispage{\baselineskip}\textit{Introduction.}---Dipole emitters such as molecules, quantum dots and nano-antennas represent fundamental building blocks in various nano-optical experiments~\cite{Michler2000,Novotny2006,Weisenburger2014a,Rotenberg2014}. In general, the emission characteristics of such dipoles depend on the relative phases and amplitudes of the three electric and/or magnetic dipolar components. For instance, Huygens dipoles---composed of perpendicular and in-phase electric and magnetic dipole moments---exhibit strongly directional emission patterns, which find applications in nanoscopic localization~\cite{Neugebauer2016} and dielectric meta-surfaces~\cite{Decker2015}. As another example, spinning electric or magnetic dipoles in proximity to an optically denser medium also exhibit directional emission and couple directionally to guided modes~\cite{Aiello2015,Bliokh2015}. Furthermore, the far field of spinning dipoles in free space is split into two half-spaces with opposite signs of helicity~\cite{OConnor2014}, an effect known as the giant spin Hall effect of light~\cite{Rodriguez-Herrera2010}. The extent of the far-field spin separation is thereby linked to the ellipticity of the polarization of the dipole~\cite{OConnor2014,Rodriguez-Herrera2010}. 

In this letter, we explore the relation between the dipole ellipticity and the far-field spin splitting. In particular, we derive a straight-forward formalism, that allows for determining the ellipticity of the dipole moment by measuring the far-field positions and helicities of points of circular polarization (C points~\cite{Nye1983}). Additionally, we propose a technique to resolve the far-field C points in the low intensity region of highly eccentric dipoles by using a method similar to quantum weak measurements~\cite{Aharonov1988,Duck1989}, which found application in optics for observing beam shift phenomena~\cite{Hosten2008,Dennis2012,Gorodetski2012}. Here, we theoretically show that the far field projection onto a spatially varying post-selected polarization state~\footnote{Postselection in optics is used in different contexts: It relates to correlated modes and conditioning (e.g. heralded single photon sources~\cite{Kok2007}), and to correlated properties in one mode (e.g. strong interference of orthogonally polarized fields by polarization projection~\cite{Hosten2008,Dennis2012,Gorodetski2012}). The usage here refers to the latter.} allows to create well-separated asymmetric far-field intensity patterns, which indicate the helicity and angular separation of the C points and, accordingly, the dipole ellipticity. Finally, we demonstrate an experimental implementation of the scheme and determine the ellipticity of the dipole moment induced in a scatterer on a dielectric interface.

\begin{figure} 
  \includegraphics[width=0.48\textwidth]{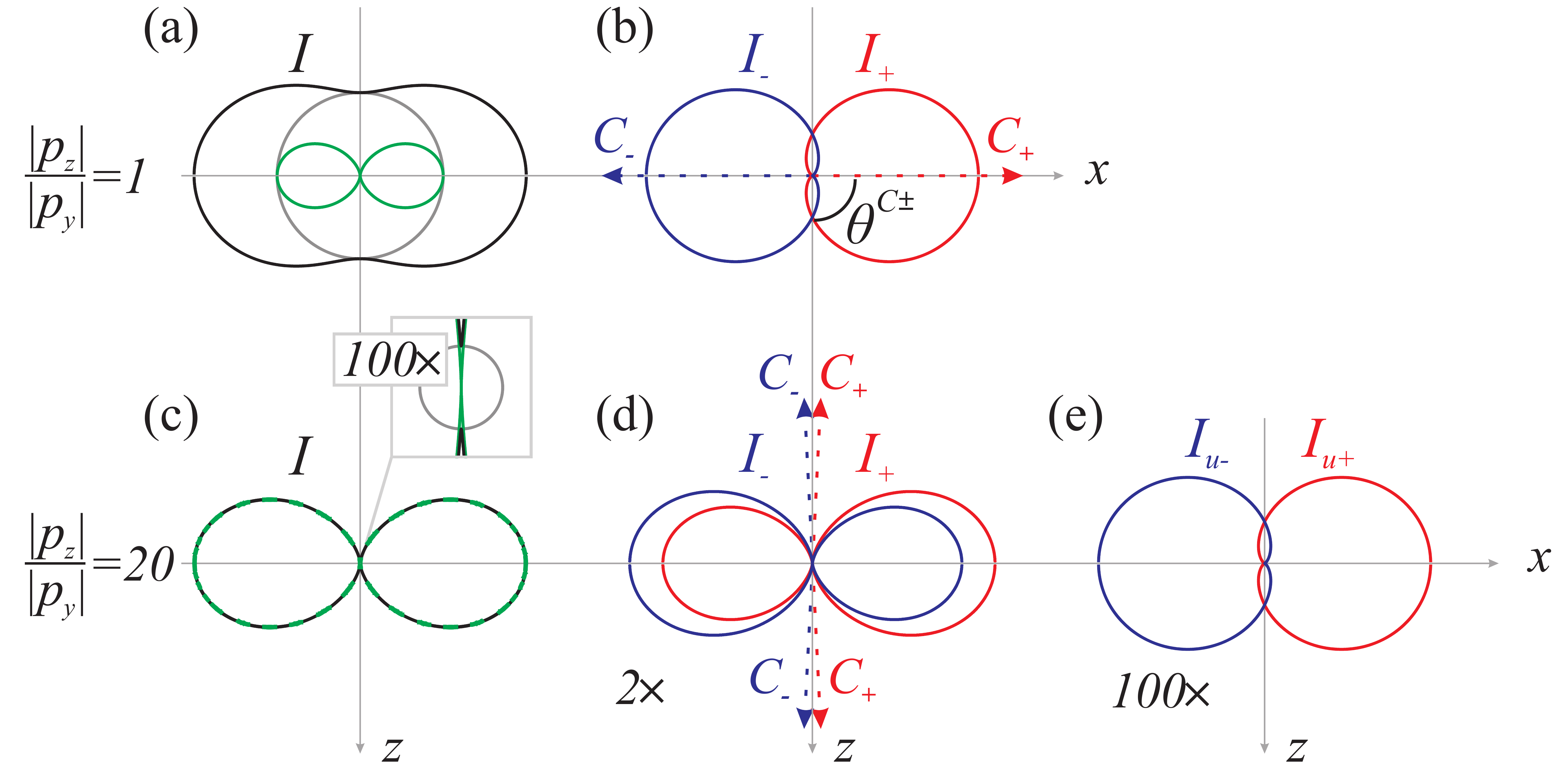}
  \caption{Free-space far-field radiation of dipoles spinning around the $x$-axis. The upper row corresponds to a circular polarized dipole, $\left|p_{z}\right|/\left|p_{y}\right|=1$, with (a) showing the total far-field intensity $I$ (black line), $I_{TM}$ of $p_z$ (green line), and $I_{TE}$ of $p_y$ (gray line). (b) depicts left-and right-handed circular polarization, $I_{-}$ and $I_{+}$. The the far-field C points ($C_{\pm}$) are highlighted by red and blue arrows. (c) and (d) show the corresponding intensities for a strongly elliptical dipole moment, $\left|p_{z}\right|/\left|p_{y}\right|=20$. The inset in (c) represents the central part magnified by a factor of $100$, and (d) is magnified by a factor of $2$. (e) represents the far field for the post-selected polarization states $I_{u-}$ and $I_{u+}$ magnified by $100$.}
  \label{fig:freespace}
\end{figure}

\textit{Elliptically polarized dipoles in free-space.}---The far-field emission pattern of an elliptically polarized dipole in free space, whose dipole moment is, without loss of generality, parallel to the $y$-$z$-plane, $\mathbf{p}=p_{y}\mathbf{e}_{y}+p_{z}\mathbf{e}_{z}\equiv \left|p_{y}\right|\mathbf{e}_{y}+\operatorname{exp}\left(\imath \Delta\varphi\right)\left|p_{z}\right|\mathbf{e}_{z}$, with $\Delta\varphi$ the relative phase between the dipole components, is given by~\cite{Jackson1999,Novotny2006}
\begin{align}\label{eqn:ff0a}
\mathbf{E}=\left(\begin{matrix}
E_{TE} \\
E_{TM}
\end{matrix}\right)
\propto \hat{M}\mathbf{p} 
=
\left(\begin{matrix}
\frac{k_{x}}{k_{\bot}}&0\\
\frac{k_{y}k_{z}}{k_{\bot}k_{0}}&-\frac{k_{\bot}}{k_{0}}
\end{matrix}\right)
\left(\begin{matrix}
p_{y} \\
p_{z}
\end{matrix}\right)
\text{,}
\end{align}
where $k_{\bot}=\left(k_{x}^2+k_{y}^2\right)^{1/2}$, $k_{z}=\pm \left(k_{0}^2-k_{\bot}^2\right)^{1/2}$, and the sign of $k_{z}$ depends on the half-space ($z\gtrless 0$). TE and TM indicate transverse electric and transverse magnetic far-field components and the matrix $\hat{M}\left(k_x,k_y\right)$ describes the overlap between the angular spectrum and the dipole moment~\cite{Novotny2006}. In the $x$-$z$-plane ($k_{y}=0$), the matrix becomes diagonal:
\begin{align}\label{eqn:ff0b}
\hat{M}\left(k_x,0\right)&=
\left(\begin{matrix}
\frac{k_{x}}{\left|k_{x}\right|}&0\\
0&-\frac{\left|k_{x}\right|}{k_{0}}
\end{matrix}\right)
\text{.}
\end{align}
Figure~\ref{fig:freespace}(a) indicates the far-field intensity $I\propto\left|\mathbf{E}\right|^{2}$ of a circularly polarized dipole ($\Delta\varphi=\pi/2$ and $\left|p_{z}\right|=\left|p_{y}\right|$) as a black line in the $x$-$z$-plane. 
The emission patterns of the individual components are plotted in green ($p_{z}$) and gray ($p_{y}$). Within the chosen plane of observation, the far field of $p_{z}$ is TM (in-plane) polarized, while the far field of $p_{y}$ is TE (out-of-plane) polarized. The relative phase between the components of the circular polarized dipole is preserved in the relative phase between $E_{TE}$ and $E_{TM}$, leading to left- and right-handed circular polarization in the far field. The corresponding circularly polarized intensity components, $I_{\pm}\propto\left|E_{TE}\mp \imath E_{TM}\right|^2$, are plotted in red and blue in Fig.~\ref{fig:freespace}(b), showing strongly directional far-field patterns with respect to the $z$-axis. Owing to the phase preservation, the C points are found by requiring $\left|E_{TE}\right|=\left|E_{TM}\right|$, which is represented graphically by the tangent points of the green and gray curves in Fig.~\ref{fig:freespace}(a). From Eqs.~\eqref{eqn:ff0a}~and~\eqref{eqn:ff0b}, the angles of the far-field C points can be determined by
\begin{align}\label{eqn:ff0c}
k^{C\pm}_{x}=\pm\left|\frac{p_{y}}{p_{z}}\right|k_{0}
\text{.}
\end{align}
\enlargethispage{\baselineskip}For $\left|p_{z}\right|=\left|p_{y}\right|$, we obtain $k^{C\pm}_{x}=\pm k_{0}$, implying that the two C points of opposite helicity---$C_{+}$ and $C_{-}$ highlighted by the red and blue dashed arrows in Fig.~\ref{fig:freespace}(b)---occur exactly on the $x$-axis and the opening angle between the C points and the $z$-axis is defined by $\theta^{C\pm}=\left|\operatorname{sin}^{-1}\left(k^{C\pm}_{x}/k_{0}\right)\right|=\pi/2$. Hence, the two C points are on opposite sides of the $z$-axis and their visibility is maximized.

In contrast, when we consider an elliptically polarized dipole moment with $\left|p_y\right|<\left|p_z\right|$, we change the weightings of the TE- and TM-polarized far fields of $p_y$ and $p_z$. As an example, we plot the far-field intensity (solid black line), which resembles the shape of the TM-polarized component (green line), for a highly eccentric dipole with an amplitude ratio of $\left|p_z\right|/\left|p_y\right|=20$ in Fig.~\ref{fig:freespace}(c). The magnified inset additionally shows the TE-polarized component indicated in gray. Instead of two tangency points, we now find four crossing points---this is four C points---at which the amplitudes of TE- and TM-polarized far fields match. Again, we mark $C_{\pm}$ as red and blue arrows in the cross-sections of $I_{\pm}$ in Fig.~\ref{fig:freespace}(d). The circular polarization components exhibit---with respect to the $z$-axis---a much weaker directionality in comparison to the circular polarized dipole in Fig.~\ref{fig:freespace}(b). Especially, the angle between the C points and the $z$-axis is small, with $\theta^{C\pm}\approx 1/20$.  

In the limit of a highly eccentric dipole $\left|p_y\right|\ll\left|p_z\right|$ the $k$-vectors of the C points are almost aligned with the $z$-axis, hidden in the low intensity region of the emission [see Fig.~\ref{fig:freespace}(c) and (d)], originating from the zero emission of $p_z$ and the relatively weak emission of $p_y$. However, it is possible to resolve the C points in the low intensity region, by using a method inspired by quantum weak measurements~~\cite{Aharonov1988,Dressel2015,Hosten2008}. For that purpose, we project the far field onto a polarization state almost orthogonal to the TM-polarization created by $p_z$, favoring the TE-polarized emission of the weak $p_y$. The actual post-selection polarization state $\mathbf{u}_{\pm}=\left(u_{TE},\pm u_{TM}\right)$ is optimized by choosing $u_{TE}=\left|p_{z}\right|/\left|\mathbf{p}\right|$ and $u_{TM}=\imath\left|p_{y}\right|/\left|\mathbf{p}\right|$, effectively compensating the amplitude difference between $p_y$ and $p_z$ in Eqs.~\eqref{eqn:ff0a}~and~\eqref{eqn:ff0b}. For the intensity pattern of the projected polarization state we obtain
\begin{align}\label{eqn:ff0d}
I_{u \pm}\propto \left|\mathbf{E}\mathbf{u}_{\pm}^{*}\right|^{2}=
\left|\frac{k_{x}}{\left|k_{x}\right|} \pm\imath\frac{\left|k_{x}\right|}{k_{0}}\right|^{2}\frac{\left|p_{y}\right|\left|p_{z}\right|}{\left|\mathbf{p}\right|^{2}}
\text{,}
\end{align}
resulting in the strongly asymmetric intensity cross-sections as they are shown for the example of $\left|p_z\right|/\left|p_y\right|=20$ in Fig.~\ref{fig:freespace}(e). The far-field intensities are identical in shape with respect to the left- and right-handed circular polarization patterns of the circular polarized dipole in Fig.~\ref{fig:freespace}(b), which means the visibility is maximized, although the overall intensity is reduced by two orders of magnitude in comparison to $I_{\pm}$ in Fig.~\ref{fig:freespace}(d). However, and most importantly, the angular separation $\Delta k^{C\pm}$ and, therefore, the dipole moment ratio [see Eq.~\eqref{eqn:ff0c}] can be deduced from the directionality of the post-selected polarization state. As a next step, with the goal of an experimental implementation, the scheme can be adapted for the case of a dipole in close proximity to a glass substrate.

\textit{Elliptically polarized dipoles on interface.}---We consider a dipole situated in air ($z<0$), with distance $d$ to an optically denser medium (glass with refractive index $n=1.5$, $z>0$). For this situation, the backward emission into the air half-space is suppressed with respect to the emission into the optically denser glass~\cite{Lukosz1977b}. Hence, we describe the emission in forward direction ($z>0$). Considering only $p_y$ and $p_z$, the far-field intensity pattern $I\left(k_x,k_y\right)$ emitted into the glass half-space $z>0$ is~\cite{Novotny2006}
\begin{align}\label{eqn:ff1}
&I\left(k_x,k_y\right)\propto \left|
\left(\begin{matrix}
E_{TE} \\
E_{TM}
\end{matrix}\right)
\right|^{2}\propto \left|A\hat{T}\hat{M}\mathbf{p}\right|^{2}\text{,}\\\label{eqn:ff2}
A&=\sqrt{k_{0}^{2}n^{2}-k_{\bot}^{2}}/k_{z}\cdot e^{\imath k_{z}d}\text{,}\quad
\hat{T}=
\left(\begin{matrix}
t_{TE}&0\\
0&t_{TM}
\end{matrix}\right)\text{.}
\end{align}
The transmission matrix $\hat{T}\left(k_{\bot}\right)$ contains the Fresnel coefficients $t_{TE}$ and $t_{TM}$, and $A\left(k_{\bot}\right)$, which depends on the distance $d$ between the dipole and the interface, is required for energy conservation~\cite{Novotny2006}. Similar to the discussion of the dipole in free space, we consider the emission within the $x$-$z$-plane and a spinning dipole with phase difference $\Delta\varphi=\pi/2$ between $p_y$ and $p_z$. In order to observe far-field C points, two conditions need to be fulfilled. Firstly, $\Delta\varphi$ needs to be preserved in the far-field components $E_{TE}$ and $E_{TM}$. This only holds true below the critical angle defined by $k_{x}=k_{0}$, since $t_{TE}$ and $t_{TM}$ are real for $k_{x}\leq k_{0}$ but complex and with different phase retardations for $k_{x}>k_{0}$. Secondly, the amplitudes of both field components need to be equal, $\left|E_{TE}\right|=\left|E_{TM}\right|$. From Eqs.~\eqref{eqn:ff1}-\eqref{eqn:ff2} it follows that
\begin{align}\label{eqn:ff3}
\left|\frac{t_{TE}p_{y}}{k_{x}} \right|=\left|\frac{t_{TM}p_{z}}{k_{0}}\right|\text{.}
\end{align}
By solving Eq.~\eqref{eqn:ff3} we can calculate $k^{C\pm}_{x}$. For $\left|p_y\right|\ll\left|p_z\right|$, we obtain the simplified expression, $k_{x}^{C\pm}\approx\pm\left|p_{y} \right|/ \left|{p_{z}}\right| k_{0}$, resembling Eq.~\eqref{eqn:ff0c}. Accordingly, the angular separation is given by $\theta^{C\pm}=\operatorname{sin}^{-1}\left(k_{x}^{C\pm}/nk_{0}\right)\approx \left|p_{y}\right|/\left|p_{z}n\right|$. 

Besides the angular separation between the C points, the intensity ratio $R$ between the maximum intensity $I_{\text{max}}$, which in the chosen geometry occurs at the the critical angle ($k_{x}=k_{0}$), and the intensity at the C points ($k_{x}=k_{x}^{C\pm}$) provides a practical measure to indicate the visibility of the C points:
\begin{align}\label{eqn:ff8}
R=\frac{I\left(k_{x}^{C\pm},0\right)}{I_{\text{max}}}\text{.}
\end{align}
For $\left|p_y\right|\ll\left|p_z\right|$, we result in $R\approx 2\left|p_{y}\right|^{2}/\left(1+n\right)^{2}\left|p_{z}\right|^{2}$. Hence, only a small fraction of the light is emitted into the angular region containing the C points. The combination of the low visibility scaling with $R\propto\left|p_{y}\right|^{2}/\left|p_{z}\right|^{2}$ and the small angular distance between the C points scaling with $\theta^{C\pm}\propto\left|p_{y}\right|/\left|p_{z}\right|$ would make it virtually impossible to experimentally measure the C points of highly eccentric dipoles. 

\begin{figure} 
  \includegraphics[width=0.48\textwidth]{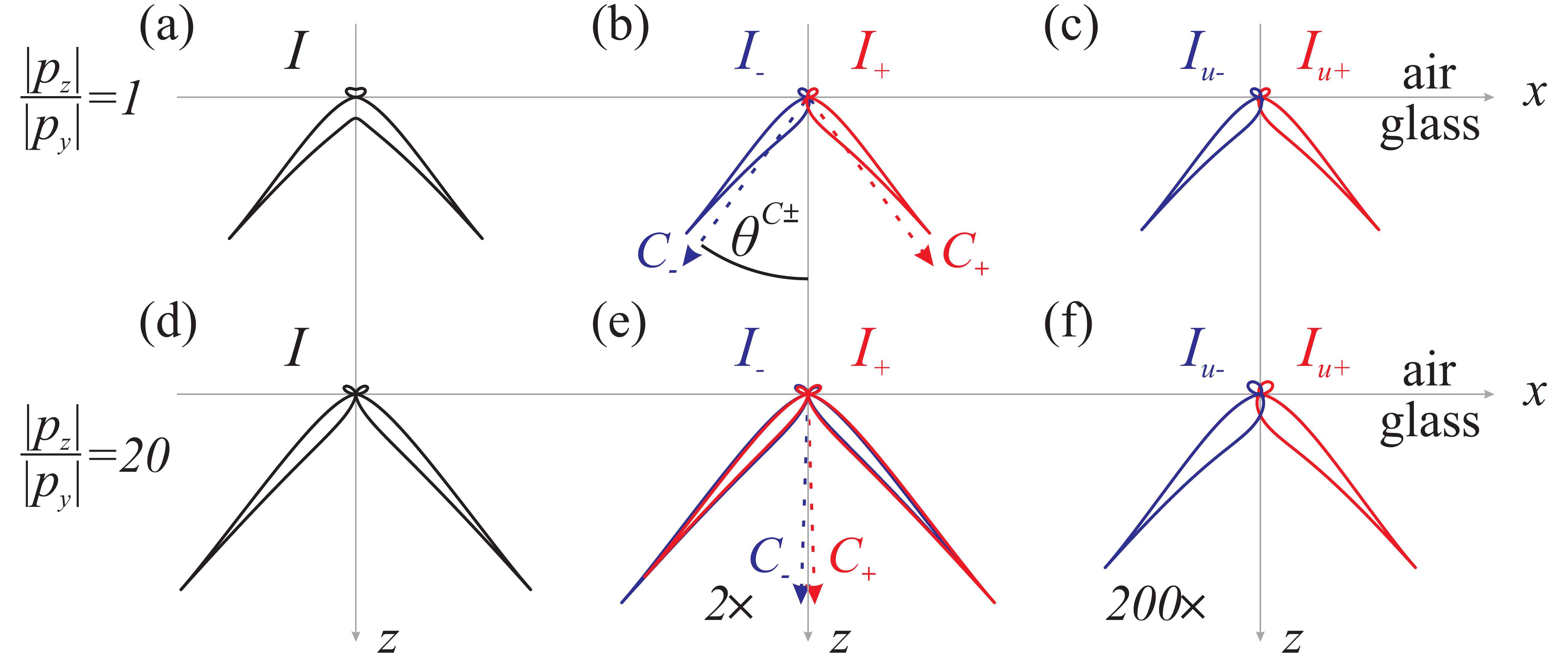}
  \caption{Far-field radiation of dipoles spinning around the $x$-axis above a dielectric air-glass interface. (a) and (b) show $I$, $I_{-}$ and $I_{+}$ for a circular polarized dipole, with red and blue arrows highlighting $C_{\pm}$. (c) represents the corresponding post-selected polarization states, $I_{u-}$ and $I_{u+}$, optimized to generate perfect directionality ($1:0$) on the critical angle. (d)-(f) show corresponding far-field intensity patterns for a strongly elliptical dipole moment, $\left|p_{z}\right|/\left|p_{y}\right|=20$.}
  \label{fig:interface}
\end{figure}

\enlargethispage{\baselineskip}To visualize our findings, we exemplarily depict the far-field intensity cross-sections for a circular polarized dipole ($\left|p_z\right|/\left|p_y\right|=1$) and a highly eccentric dipole ($\left|p_z\right|/\left|p_y\right|=20$) in Figs.~\ref{fig:interface}(a)~and~(d), for a wavelength of $\lambda=530\text{ nm}$ and a distance of $d=40\text{ nm}$.
The corresponding circular polarization components plotted in Figs.~\ref{fig:interface}(b)~and~(e) exhibit a very strong directionality for $\left|p_z\right|/\left|p_y\right|=1$ and a much weaker directionality for $\left|p_z\right|/\left|p_y\right|=20$. For the circularly polarized dipole, two C points occur below but close to the critical angle at $\theta^{C\pm} \approx 0.20\pi$ with $R \approx 0.33$. In the upper half-space ($z<0$), no C points occur due to the interference of the direct emission with the light reflected at the interface, which changes the relative phase between $E_{TE}$ and $E_{TM}$. In contrast to the circularly polarized dipole, the two C points of the highly eccentric dipole appear close to the $z$-axis. Similar to the elliptically polarized dipole in free-space [see Figs.~\ref{fig:freespace}(d)-(f)], $C_{+}$ and $C_{-}$ are not only barely separated in $k$-space ($\theta^{C\pm}\approx 0.01\pi$), but also hidden in the low intensity region ($R\approx 1.5\cdot 10^{-3}$). However, the visibility can be enhanced by choosing an appropriate post-selected polarization state,
\begin{align}\label{eqn:ff9}
\mathbf{u}_{\pm}=
\left(\begin{matrix}
u_{TE}\\
\pm u_{TM}
\end{matrix}\right)\propto
\left(\begin{matrix}
t_{TM}\left(k_{0}\right)\left|p_{z}\right|\\
\pm \imath t_{TE}\left(k_{0}\right)\left|p_{y}\right|
\end{matrix}\right)\text{,}
\end{align}
which compensates the amplitude difference of the far fields of $p_{y}$ and $p_{z}$ at the critical angle ($k_{x}=k_{0}$). By applying this scheme to the far fields of the circular and the highly eccentric dipole moment, we obtain the far-field intensities, $I_{u \pm}\propto\left|\mathbf{E}\mathbf{u}_{\pm}^{*}\right|^{2}$, plotted in Figs.~\ref{fig:interface}(c)~and~(f). In particular for the elliptical dipole moment, the visibility is strongly enhanced with respect to the projection onto the circular polarization basis shown in Fig.~\ref{fig:interface}(e). Since the maximum of $I_{u \pm}$ is at the same angular position (critical angle) as the zero value of $I_{u \mp}$, the visibility is maximized. However, the increased visibility is reached at the cost of the overall intensity being reduced by two orders of magnitude. Nevertheless, we can use the described approach to measure hardly separated C points. It is worth noting here already that this method measuring the C point splitting using a weak-measurement-inspired approach will also enable the precise experimental retrieval of the dipole moments themselves. The details will be explained with the aid of an experimental example. 

\begin{figure} 
  \includegraphics[width=0.45\textwidth]{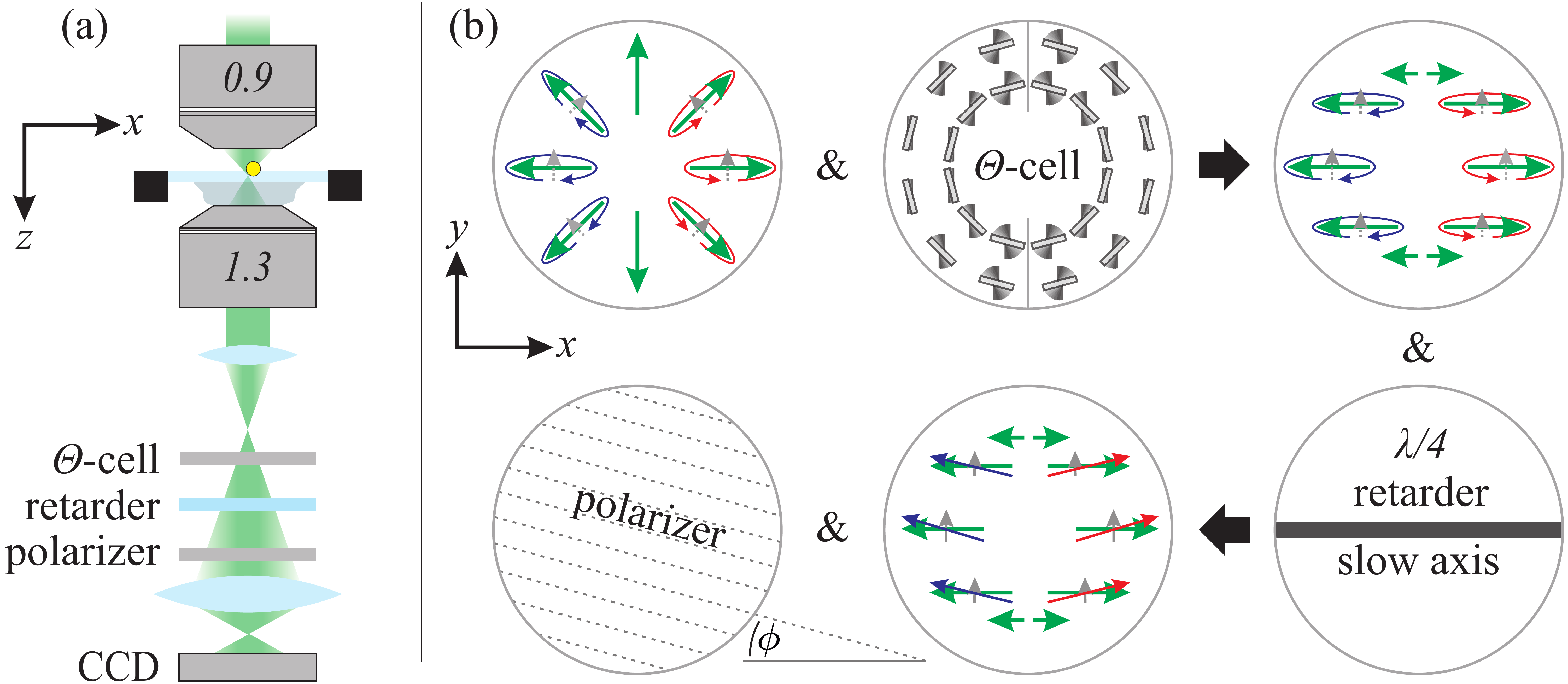}
  \caption{Experimental setup and measurement scheme. (a) sketch of the experimental setup. (b) Projection onto the polarization states $\mathbf{u}_{\pm}=\left[\cos\left(\phi\right),\imath\sin\left(\phi\right)\right]$, with $\pm$ indicating $\phi\gtrless 0$.}
  \label{fig:setup}
\end{figure}

The utilized measurement setup is sketched in Fig.~\ref{fig:setup}(a).
An incoming radially polarized beam ($\lambda=530\text{ nm}$, beam width $w_{0}=1.1\text{ mm}$) focused by a microscope objective with a numerical aperture ($\text{NA}$) of 0.9 and focal length $f=2.0\text{ mm}$ excites a gold nano-sphere with a diameter of $80\text{\,nm}$ sitting on a glass substrate. Because of the small size, the particle can be approximated as a dipole-like scatterer with its dipole moment being proportional to the local excitation field, $\mathbf{p}\propto \mathbf{E}$~\cite{Lee2007}. When the particle is in the center of the focal spot, we induce a $p_{z}$ dipole moment~\cite{Neugebauer2014}. However, a sub-wavelength shift of the particle along the $y$-direction induces an additional weak $p_{y}$ dipole component, due to the non-zero transverse field component at such an off-center position~\cite{Neugebauer2014}. The relative phase between $p_z$ and $p_y$ is close to $\pm\pi/2$, because of the transverse spin arising in tightly focused radially polarized beams~\cite{Neugebauer2014,Neugebauer2015}. The ratio of the dipole amplitudes is controlled by the distance between the particle and the center of the focal spot~\cite{Neugebauer2014}. The scattered light emitted into the glass substrate is collected by a second objective (immersion type, $\text{NA}=1.3$) in confocal configuration. Behind the objective, we perform the projection of the polarization state of the scattered light onto the desired polarization state $\mathbf{u}_{\pm}$. At first, we image the back focal plane of the lower objective onto a $\Theta$-cell. Such a $\Theta$-cell contains liquid crystals, which locally rotate the incoming polarization state~\cite{Stalder1996a}, effectively converting $E_{TE}$ and $E_{TM}$ to $E_{y}$ and $E_{x}$. Additionally, the $\Theta$-cell introduces a small phase shift between $E_{TE}$ and $E_{TM}$, which can be neglected in this proof-of-principle experiment. For illustration, the effect of the $\Theta$-cell on an incoming polarization state is sketched in Fig.~\ref{fig:setup}(b). In front of the $\Theta$-cell, the light is mainly TM-polarized (green arrows). However, a weak TE-component (gray arrows) is present as well. The TM-component is symmetric with respect to the optical axis of the system ($z$-axis) and the TE-component is anti-symmetric with respect to the $y$-axis and phase-shifted with respect to the TM-component by $\pm \pi/2$. This exemplary polarization state is chosen for this discussion, since it is similar to the expected left- and right-handed elliptically polarized far fields (see red and and blue arrows) of a transversely spinning dipole~\cite{Rodriguez-Herrera2010}. The local polarization rotation of the $\Theta$-cell is sketched in gray. The polarization distribution behind the $\Theta$-cell is sketched on the right, with the TM-component (TE-component) being converted into $x$-polarization ($y$-polarization). By introducing a $\lambda/4$-retarder, we can further transform the locally elliptical polarization distribution into purely linearly polarization with varying orientation of the polarization axis. Utilizing a linear polarizer set to an angle $\phi$ with respect to the $x$-axis, we realize the projection onto the polarization state $\mathbf{u}_{\pm}=\left[\cos\left(\phi\right),\imath\sin\left(\phi\right)\right]$, where the index $\pm$ indicates the positive or negative sign of $\phi$. 

\begin{figure} 
  \includegraphics[width=0.48\textwidth]{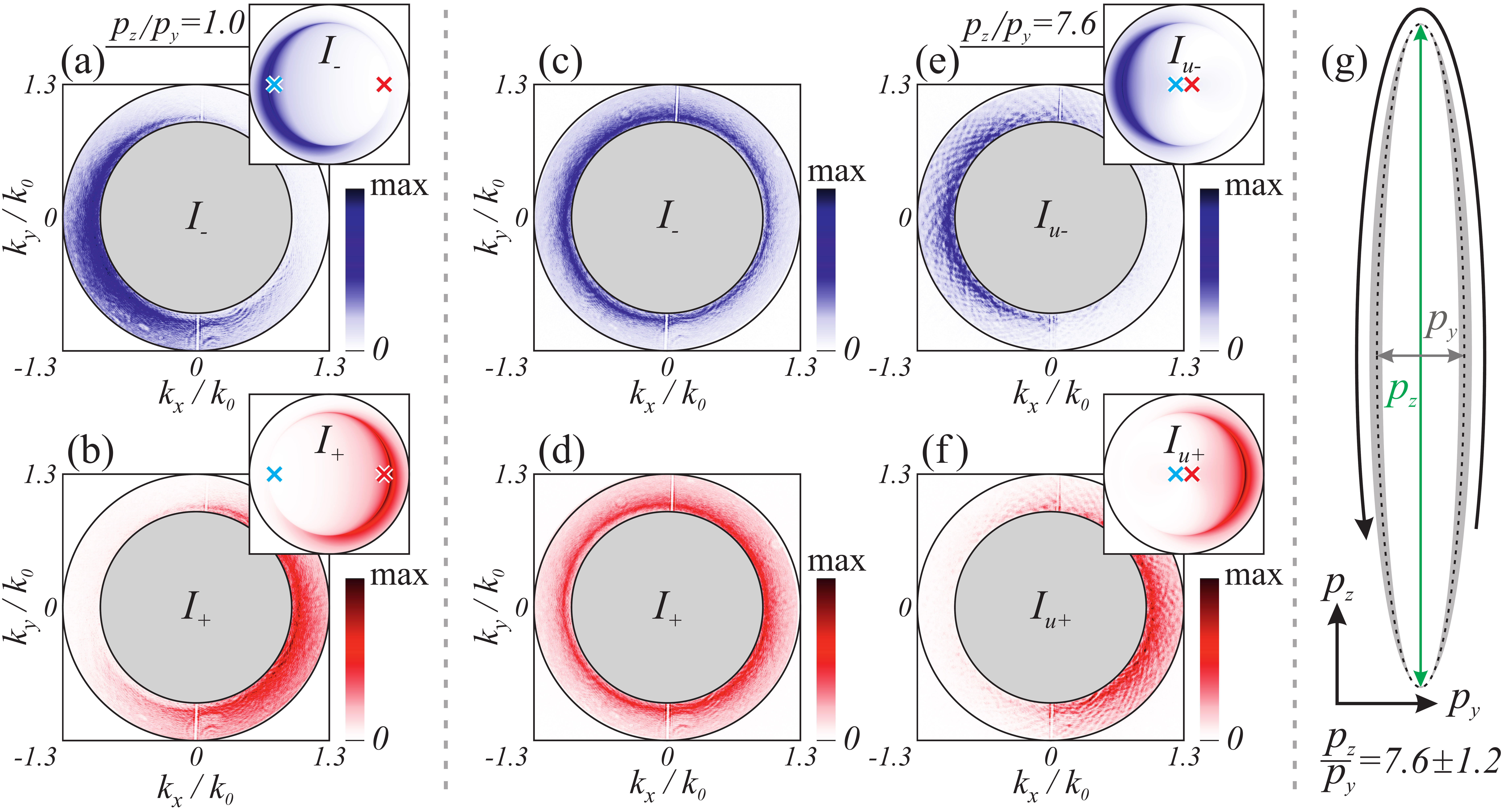}
  \caption{Polarization-resolved back focal plane (BFP) images. (a) and (b) show right- and left-handed circular polarization distributions (normalized to their common maximum amplitude, depicted angular range defined by $0.95 \leq k_{\bot}/k_{0}\leq 1.3$) for a particle position of $x\approx 0\text{\,nm}$ and $y\approx 200\text{\,nm}$, where $\left|p_{z}\right|/\left|p_{y}\right|\approx 1$. Corresponding calculated BFP images are shown as insets (red and blue crosses indicate the C points). (c) and (d) show circular polarization distributions for a particle position of $x\approx 0\text{\,nm}$ and $y\approx 30\text{\,nm}$. (e) and (f) show BFP images for the same position as (c) and (d), but for a post-selected polarization state of $\mathbf{u}_{\pm}=\left[\imath\cos\left(\phi\right),\sin\left(\phi\right)\right]$ with $\phi=\pm 5^{\circ}$. Theoretically calculated BFP images including the reconstructed C points are shown as insets. (g) The dashed black line indicates the reconstructed polarization ellipse, with error margins, spin and dipole moments indicated as gray background, black vector and green ($p_{z}$) and gray ($p_{y}$) vectors, respectively.}
  \label{fig:bfp}
\end{figure}

\enlargethispage{\baselineskip}Finally, we experimentally investigate the far-field emission patterns polarization resolved for two different particle positions. In Figs.~\ref{fig:bfp}(a)~and~(b), we depict the left- and right-handed circularly polarized far fields $I_{-}$ and $I_{+}$ within the angular range defined by $0.95 \leq k_{\bot}/k_{0}\leq 1.3$ for a particle position of $x\approx 0\text{\,nm}$ and $y\approx 200\text{\,nm}$, where we expect the induced dipole moment to be close to circular~\cite{Neugebauer2014}. 
For comparison, we depict the theoretical distributions of $I_{-}$ and $I_{+}$ as insets, calculated for $\Delta\varphi=\pi/2$ and $\left|p_y\right|=\left|p_z\right|$. As mentioned above, the C points occur below but close to the critical angle (see red and blue crosses). Experiment and theory are in very good agreement, indicating that the experimentally excited dipole closely resembles the theoretically assumed dipole moment.\\
As a second position, we choose the particle to be much closer to the optical axis, with $x\approx 0\text{\,nm}$ and $y\approx 30\text{\,nm}$, obtaining the left- and right-handed circularly polarized far fields as they are depicted in Figs.~\ref{fig:bfp}(c)~and~(d). Because of the smaller distance to the optical axis, the longitudinal electric dipole component $p_{z}$ dominates the transverse one ($p_{y}$), and in comparison to Figs.~\ref{fig:bfp}(a)~and~(b), the spin splitting is much weaker. It becomes clear that the C points are hard to observe, since they are barely separated in the low intensity of the far-field pattern. More importantly, they are in the angular region far below $\text{NA}=0.9$, where the scattered light is actually interfering with the transmitted excitation beam. However, it is still possible to determine the position of these C points of the scattered light by adjusting the angle of the polarizer to reach maximum visibility at the critical angle. Here, we obtain the strongly directional far-field patterns in Figs.~\ref{fig:bfp}(e)~and~(f) for a polarizer angle of $\phi \approx 5^{\circ}$. By comparing the experimental post-selection polarization state $\mathbf{u}_{\pm}=\left[\cos\left(5^{\circ}\right),\pm\imath\sin\left(5^{\circ}\right)\right]$ with Eq.~\eqref{eqn:ff9} we can determine the dipole moment amplitude ratio, $\left|p_{z}\right|/\left|p_{y}\right|\approx 7.6\pm 1.2$, and the sign of the spin, here pointing in positive $x$-direction~\cite{Neugebauer2015}. The result is illustrated as polarization ellipse (dashed black line) in Fig.~\ref{fig:bfp}(g), where the margin of error is indicated in the background as gray area, the long and short axis of the dipole are shown as green and gray vectors, and the spinning direction is sketched as black vector. In particular, the good overlap between experiment and theoretically calculated insets in Figs.~\ref{fig:bfp}(e)~and~(f)---where we also depict the angular positions of the C points for the sake of completeness---indirectly validates the experimental post-selection technique.

\textit{Discussion and Conclusion.}---We investigated the far-field spin-splitting of elliptically polarized dipoles in free space. Careful examination of the dipole emission patterns revealed two pairs of C points, with the angular separation of each pair scaling with the ratio of the minor and major axes of the dipole moment. Utilizing TE (azimuthal) and TM (radial) polarization states as a new basis in optical weak measurements, we experimentally and theoretically demonstrated that by projecting the far field onto a spatially varying post-selected polarization state, we are able to reveal the angular separation and the helicities of these points even for highly eccentric dipole moments. Most importantly, we can extract the direction of spin and the ratio of the minor and major axes of the dipole moment from such measurements, since the position and handedness of the C points entail information on the polarization state of the dipole. Hence, the presented novel approach allows for a precise experimental retrieval of the complex dipole moments induced in nanoparticles, extending the current arsenal of techniques for measuring orientation and spin of individual dipoles~\cite{Lieb2004,Foreman2008}. Our weak measurement scheme is particularly useful for measuring small polarization deviations in comparison to an ideal dipole polarization state (here $p_{z}$). For the experimental demonstration, we adapted the approach for dipoles close to a dielectric interface, and measured the ellipticity of the dipole moment induced by a tightly focused beam.\\
Although we considered only elliptically polarized electric dipole moments parallel to the $y$-$z$-plane, the approach can also be adopted for investigating more complex dipole and multipole combinations. Ultimately, the far-field polarization projection might find application in high-precision nano-metrology and microscopy. 

\begin{acknowledgments}
We gratefully acknowledge fruitful discussions with Yuri Gorodetski.
\end{acknowledgments}
%
\bibliography{bib}

\begin{thebibliography}{27}%
\makeatletter
\providecommand \@ifxundefined [1]{%
 \@ifx{#1\undefined}
}%
\providecommand \@ifnum [1]{%
 \ifnum #1\expandafter \@firstoftwo
 \else \expandafter \@secondoftwo
 \fi
}%
\providecommand \@ifx [1]{%
 \ifx #1\expandafter \@firstoftwo
 \else \expandafter \@secondoftwo
 \fi
}%
\providecommand \natexlab [1]{#1}%
\providecommand \enquote  [1]{``#1''}%
\providecommand \bibnamefont  [1]{#1}%
\providecommand \bibfnamefont [1]{#1}%
\providecommand \citenamefont [1]{#1}%
\providecommand \href@noop [0]{\@secondoftwo}%
\providecommand \href [0]{\begingroup \@sanitize@url \@href}%
\providecommand \@href[1]{\@@startlink{#1}\@@href}%
\providecommand \@@href[1]{\endgroup#1\@@endlink}%
\providecommand \@sanitize@url [0]{\catcode `\\12\catcode `\$12\catcode
  `\&12\catcode `\#12\catcode `\^12\catcode `\_12\catcode `\%12\relax}%
\providecommand \@@startlink[1]{}%
\providecommand \@@endlink[0]{}%
\providecommand \url  [0]{\begingroup\@sanitize@url \@url }%
\providecommand \@url [1]{\endgroup\@href {#1}{\urlprefix }}%
\providecommand \urlprefix  [0]{URL }%
\providecommand \Eprint [0]{\href }%
\providecommand \doibase [0]{http://dx.doi.org/}%
\providecommand \selectlanguage [0]{\@gobble}%
\providecommand \bibinfo  [0]{\@secondoftwo}%
\providecommand \bibfield  [0]{\@secondoftwo}%
\providecommand \translation [1]{[#1]}%
\providecommand \BibitemOpen [0]{}%
\providecommand \bibitemStop [0]{}%
\providecommand \bibitemNoStop [0]{.\EOS\space}%
\providecommand \EOS [0]{\spacefactor3000\relax}%
\providecommand \BibitemShut  [1]{\csname bibitem#1\endcsname}%
\let\auto@bib@innerbib\@empty
\bibitem [{\citenamefont {Michler}\ \emph {et~al.}(2000)\citenamefont
  {Michler}, \citenamefont {Kiraz}, \citenamefont {Becher}, \citenamefont
  {Schoenfeld}, \citenamefont {Petroff}, \citenamefont {Zhang}, \citenamefont
  {Hu},\ and\ \citenamefont {Imamoglu}}]{Michler2000}%
  \BibitemOpen
  \bibfield  {author} {\bibinfo {author} {\bibfnamefont {P.}~\bibnamefont
  {Michler}}, \bibinfo {author} {\bibfnamefont {A.}~\bibnamefont {Kiraz}},
  \bibinfo {author} {\bibfnamefont {C.}~\bibnamefont {Becher}}, \bibinfo
  {author} {\bibfnamefont {W.~V.}\ \bibnamefont {Schoenfeld}}, \bibinfo
  {author} {\bibfnamefont {P.~M.}\ \bibnamefont {Petroff}}, \bibinfo {author}
  {\bibfnamefont {L.}~\bibnamefont {Zhang}}, \bibinfo {author} {\bibfnamefont
  {E.}~\bibnamefont {Hu}}, \ and\ \bibinfo {author} {\bibfnamefont
  {A.}~\bibnamefont {Imamoglu}},\ }\href {\doibase
  10.1126/science.290.5500.2282} {\bibfield  {journal} {\bibinfo  {journal}
  {Science}\ }\textbf {\bibinfo {volume} {290}},\ \bibinfo {pages} {2282}
  (\bibinfo {year} {2000})}\BibitemShut {NoStop}%
\bibitem [{\citenamefont {Novotny}\ and\ \citenamefont
  {Hecht}(2006)}]{Novotny2006}%
  \BibitemOpen
  \bibfield  {author} {\bibinfo {author} {\bibfnamefont {L.}~\bibnamefont
  {Novotny}}\ and\ \bibinfo {author} {\bibfnamefont {B.}~\bibnamefont
  {Hecht}},\ }\href@noop {} {\emph {\bibinfo {title} {{Principles of
  Nano-Optics}}}},\ \bibinfo {edition} {2nd}\ ed.\ (\bibinfo  {publisher}
  {Cambridge University Press},\ \bibinfo {address} {Cambridge},\ \bibinfo
  {year} {2006})\BibitemShut {NoStop}%
\bibitem [{\citenamefont {Weisenburger}\ \emph {et~al.}(2014)\citenamefont
  {Weisenburger}, \citenamefont {Jing}, \citenamefont {H{\"{a}}nni},
  \citenamefont {Reymond}, \citenamefont {Schuler}, \citenamefont {Renn},\ and\
  \citenamefont {Sandoghdar}}]{Weisenburger2014a}%
  \BibitemOpen
  \bibfield  {author} {\bibinfo {author} {\bibfnamefont {S.}~\bibnamefont
  {Weisenburger}}, \bibinfo {author} {\bibfnamefont {B.}~\bibnamefont {Jing}},
  \bibinfo {author} {\bibfnamefont {D.}~\bibnamefont {H{\"{a}}nni}}, \bibinfo
  {author} {\bibfnamefont {L.}~\bibnamefont {Reymond}}, \bibinfo {author}
  {\bibfnamefont {B.}~\bibnamefont {Schuler}}, \bibinfo {author} {\bibfnamefont
  {A.}~\bibnamefont {Renn}}, \ and\ \bibinfo {author} {\bibfnamefont
  {V.}~\bibnamefont {Sandoghdar}},\ }\href {\doibase 10.1002/cphc.201301080}
  {\bibfield  {journal} {\bibinfo  {journal} {Chemphyschem}\ }\textbf {\bibinfo
  {volume} {15}},\ \bibinfo {pages} {763} (\bibinfo {year} {2014})}\BibitemShut
  {NoStop}%
\bibitem [{\citenamefont {Rotenberg}\ and\ \citenamefont
  {Kuipers}(2014)}]{Rotenberg2014}%
  \BibitemOpen
  \bibfield  {author} {\bibinfo {author} {\bibfnamefont {N.}~\bibnamefont
  {Rotenberg}}\ and\ \bibinfo {author} {\bibfnamefont {L.}~\bibnamefont
  {Kuipers}},\ }\href {\doibase 10.1038/nphoton.2014.285} {\bibfield  {journal}
  {\bibinfo  {journal} {Nat. Photon.}\ }\textbf {\bibinfo {volume} {8}},\
  \bibinfo {pages} {919} (\bibinfo {year} {2014})}\BibitemShut {NoStop}%
\bibitem [{\citenamefont {Neugebauer}\ \emph {et~al.}(2016)\citenamefont
  {Neugebauer}, \citenamefont {Wo{\'{z}}niak}, \citenamefont {Bag},
  \citenamefont {Leuchs},\ and\ \citenamefont {Banzer}}]{Neugebauer2016}%
  \BibitemOpen
  \bibfield  {author} {\bibinfo {author} {\bibfnamefont {M.}~\bibnamefont
  {Neugebauer}}, \bibinfo {author} {\bibfnamefont {P.}~\bibnamefont
  {Wo{\'{z}}niak}}, \bibinfo {author} {\bibfnamefont {A.}~\bibnamefont {Bag}},
  \bibinfo {author} {\bibfnamefont {G.}~\bibnamefont {Leuchs}}, \ and\ \bibinfo
  {author} {\bibfnamefont {P.}~\bibnamefont {Banzer}},\ }\href {\doibase
  10.1038/ncomms11286 OPEN} {\bibfield  {journal} {\bibinfo  {journal} {Nat.
  Commun.}\ }\textbf {\bibinfo {volume} {7}},\ \bibinfo {pages} {11286}
  (\bibinfo {year} {2016})}\BibitemShut {NoStop}%
\bibitem [{\citenamefont {Decker}\ \emph {et~al.}(2015)\citenamefont {Decker},
  \citenamefont {Staude}, \citenamefont {Falkner}, \citenamefont {Dominguez},
  \citenamefont {Neshev}, \citenamefont {Brener}, \citenamefont {Pertsch},\
  and\ \citenamefont {Kivshar}}]{Decker2015}%
  \BibitemOpen
  \bibfield  {author} {\bibinfo {author} {\bibfnamefont {M.}~\bibnamefont
  {Decker}}, \bibinfo {author} {\bibfnamefont {I.}~\bibnamefont {Staude}},
  \bibinfo {author} {\bibfnamefont {M.}~\bibnamefont {Falkner}}, \bibinfo
  {author} {\bibfnamefont {J.}~\bibnamefont {Dominguez}}, \bibinfo {author}
  {\bibfnamefont {D.~N.}\ \bibnamefont {Neshev}}, \bibinfo {author}
  {\bibfnamefont {I.}~\bibnamefont {Brener}}, \bibinfo {author} {\bibfnamefont
  {T.}~\bibnamefont {Pertsch}}, \ and\ \bibinfo {author} {\bibfnamefont
  {Y.~S.}\ \bibnamefont {Kivshar}},\ }\href {\doibase 10.1002/adom.201400584}
  {\bibfield  {journal} {\bibinfo  {journal} {Adv. Optical Mater.}\ }\textbf
  {\bibinfo {volume} {3}},\ \bibinfo {pages} {813} (\bibinfo {year}
  {2015})}\BibitemShut {NoStop}%
\bibitem [{\citenamefont {Aiello}\ \emph {et~al.}(2015)\citenamefont {Aiello},
  \citenamefont {Banzer}, \citenamefont {Neugebauer},\ and\ \citenamefont
  {Leuchs}}]{Aiello2015}%
  \BibitemOpen
  \bibfield  {author} {\bibinfo {author} {\bibfnamefont {A.}~\bibnamefont
  {Aiello}}, \bibinfo {author} {\bibfnamefont {P.}~\bibnamefont {Banzer}},
  \bibinfo {author} {\bibfnamefont {M.}~\bibnamefont {Neugebauer}}, \ and\
  \bibinfo {author} {\bibfnamefont {G.}~\bibnamefont {Leuchs}},\ }\href
  {\doibase 10.1038/nphoton.2015.203} {\bibfield  {journal} {\bibinfo
  {journal} {Nat. Photon.}\ }\textbf {\bibinfo {volume} {9}},\ \bibinfo {pages}
  {789} (\bibinfo {year} {2015})}\BibitemShut {NoStop}%
\bibitem [{\citenamefont {Bliokh}\ \emph {et~al.}(2015)\citenamefont {Bliokh},
  \citenamefont {Rodr{\'{i}}guez-Fortu{\~{n}}o}, \citenamefont {Nori},\ and\
  \citenamefont {Zayats}}]{Bliokh2015}%
  \BibitemOpen
  \bibfield  {author} {\bibinfo {author} {\bibfnamefont {K.~Y.}\ \bibnamefont
  {Bliokh}}, \bibinfo {author} {\bibfnamefont {F.~J.}\ \bibnamefont
  {Rodr{\'{i}}guez-Fortu{\~{n}}o}}, \bibinfo {author} {\bibfnamefont
  {F.}~\bibnamefont {Nori}}, \ and\ \bibinfo {author} {\bibfnamefont {A.~V.}\
  \bibnamefont {Zayats}},\ }\href {\doibase 10.1038/nphoton.2015.201}
  {\bibfield  {journal} {\bibinfo  {journal} {Nat. Photon.}\ }\textbf {\bibinfo
  {volume} {9}},\ \bibinfo {pages} {796} (\bibinfo {year} {2015})}\BibitemShut
  {NoStop}%
\bibitem [{\citenamefont {O'Connor}\ \emph {et~al.}(2014)\citenamefont
  {O'Connor}, \citenamefont {Ginzburg}, \citenamefont
  {Rodr{\'{i}}guez-Fortu{\~{n}}o}, \citenamefont {Wurtz},\ and\ \citenamefont
  {Zayats}}]{OConnor2014}%
  \BibitemOpen
  \bibfield  {author} {\bibinfo {author} {\bibfnamefont {D.}~\bibnamefont
  {O'Connor}}, \bibinfo {author} {\bibfnamefont {P.}~\bibnamefont {Ginzburg}},
  \bibinfo {author} {\bibfnamefont {F.~J.}\ \bibnamefont
  {Rodr{\'{i}}guez-Fortu{\~{n}}o}}, \bibinfo {author} {\bibfnamefont {G.~A.}\
  \bibnamefont {Wurtz}}, \ and\ \bibinfo {author} {\bibfnamefont {A.~V.}\
  \bibnamefont {Zayats}},\ }\href {\doibase 10.1038/ncomms6327} {\bibfield
  {journal} {\bibinfo  {journal} {Nat. Commun.}\ }\textbf {\bibinfo {volume}
  {5}},\ \bibinfo {pages} {5327} (\bibinfo {year} {2014})}\BibitemShut
  {NoStop}%
\bibitem [{\citenamefont {Rodr{\'{i}}guez-Herrera}\ \emph
  {et~al.}(2010)\citenamefont {Rodr{\'{i}}guez-Herrera}, \citenamefont {Lara},
  \citenamefont {Bliokh}, \citenamefont {Ostrovskaya},\ and\ \citenamefont
  {Dainty}}]{Rodriguez-Herrera2010}%
  \BibitemOpen
  \bibfield  {author} {\bibinfo {author} {\bibfnamefont {O.~G.}\ \bibnamefont
  {Rodr{\'{i}}guez-Herrera}}, \bibinfo {author} {\bibfnamefont
  {D.}~\bibnamefont {Lara}}, \bibinfo {author} {\bibfnamefont {K.~Y.}\
  \bibnamefont {Bliokh}}, \bibinfo {author} {\bibfnamefont {E.~A.}\
  \bibnamefont {Ostrovskaya}}, \ and\ \bibinfo {author} {\bibfnamefont
  {C.}~\bibnamefont {Dainty}},\ }\href {\doibase
  10.1103/PhysRevLett.104.253601} {\bibfield  {journal} {\bibinfo  {journal}
  {Phys. Rev. Lett.}\ }\textbf {\bibinfo {volume} {104}},\ \bibinfo {pages}
  {253601} (\bibinfo {year} {2010})}\BibitemShut {NoStop}%
\bibitem [{\citenamefont {Nye}(1983)}]{Nye1983}%
  \BibitemOpen
  \bibfield  {author} {\bibinfo {author} {\bibfnamefont {J.~F.}\ \bibnamefont
  {Nye}},\ }\href {\doibase 10.1098/rspa.1983.0109} {\bibfield  {journal}
  {\bibinfo  {journal} {Proc. R. Soc. Lond. A}\ }\textbf {\bibinfo {volume}
  {389}},\ \bibinfo {pages} {279} (\bibinfo {year} {1983})}\BibitemShut
  {NoStop}%
\bibitem [{\citenamefont {Aharonov}\ \emph {et~al.}(1988)\citenamefont
  {Aharonov}, \citenamefont {Albert},\ and\ \citenamefont
  {Vaidman}}]{Aharonov1988}%
  \BibitemOpen
  \bibfield  {author} {\bibinfo {author} {\bibfnamefont {Y.}~\bibnamefont
  {Aharonov}}, \bibinfo {author} {\bibfnamefont {D.}~\bibnamefont {Albert}}, \
  and\ \bibinfo {author} {\bibfnamefont {L.}~\bibnamefont {Vaidman}},\ }\href
  {\doibase 10.1103/PhysRevLett.60.1351} {\bibfield  {journal} {\bibinfo
  {journal} {Phys. Rev. Lett.}\ }\textbf {\bibinfo {volume} {60}},\ \bibinfo
  {pages} {1351} (\bibinfo {year} {1988})}\BibitemShut {NoStop}%
\bibitem [{\citenamefont {Duck}\ \emph {et~al.}(1989)\citenamefont {Duck},
  \citenamefont {Stevenson},\ and\ \citenamefont {Sudarshan}}]{Duck1989}%
  \BibitemOpen
  \bibfield  {author} {\bibinfo {author} {\bibfnamefont {I.~M.}\ \bibnamefont
  {Duck}}, \bibinfo {author} {\bibfnamefont {P.~M.}\ \bibnamefont {Stevenson}},
  \ and\ \bibinfo {author} {\bibfnamefont {E.~C.~G.}\ \bibnamefont
  {Sudarshan}},\ }\href@noop {} {\bibfield  {journal} {\bibinfo  {journal}
  {Phys. Rev. B}\ }\textbf {\bibinfo {volume} {40}},\ \bibinfo {pages} {2112}
  (\bibinfo {year} {1989})}\BibitemShut {NoStop}%
\bibitem [{\citenamefont {Hosten}\ and\ \citenamefont
  {Kwiat}(2008)}]{Hosten2008}%
  \BibitemOpen
  \bibfield  {author} {\bibinfo {author} {\bibfnamefont {O.}~\bibnamefont
  {Hosten}}\ and\ \bibinfo {author} {\bibfnamefont {P.}~\bibnamefont {Kwiat}},\
  }\href {\doibase 10.1126/science.1152697} {\bibfield  {journal} {\bibinfo
  {journal} {Science}\ }\textbf {\bibinfo {volume} {319}},\ \bibinfo {pages}
  {787} (\bibinfo {year} {2008})}\BibitemShut {NoStop}%
\bibitem [{\citenamefont {Dennis}\ and\ \citenamefont
  {G{\"{o}}tte}(2012)}]{Dennis2012}%
  \BibitemOpen
  \bibfield  {author} {\bibinfo {author} {\bibfnamefont {M.~R.}\ \bibnamefont
  {Dennis}}\ and\ \bibinfo {author} {\bibfnamefont {J.~B.}\ \bibnamefont
  {G{\"{o}}tte}},\ }\href {\doibase 10.1088/1367-2630/14/7/073013} {\bibfield
  {journal} {\bibinfo  {journal} {New Journal of Physics}\ }\textbf {\bibinfo
  {volume} {14}},\ \bibinfo {pages} {073013} (\bibinfo {year}
  {2012})}\BibitemShut {NoStop}%
\bibitem [{\citenamefont {Gorodetski}\ \emph {et~al.}(2012)\citenamefont
  {Gorodetski}, \citenamefont {Bliokh}, \citenamefont {Stein}, \citenamefont
  {Genet}, \citenamefont {Shitrit}, \citenamefont {Kleiner}, \citenamefont
  {Hasman},\ and\ \citenamefont {Ebbesen}}]{Gorodetski2012}%
  \BibitemOpen
  \bibfield  {author} {\bibinfo {author} {\bibfnamefont {Y.}~\bibnamefont
  {Gorodetski}}, \bibinfo {author} {\bibfnamefont {K.~Y.}\ \bibnamefont
  {Bliokh}}, \bibinfo {author} {\bibfnamefont {B.}~\bibnamefont {Stein}},
  \bibinfo {author} {\bibfnamefont {C.}~\bibnamefont {Genet}}, \bibinfo
  {author} {\bibfnamefont {N.}~\bibnamefont {Shitrit}}, \bibinfo {author}
  {\bibfnamefont {V.}~\bibnamefont {Kleiner}}, \bibinfo {author} {\bibfnamefont
  {E.}~\bibnamefont {Hasman}}, \ and\ \bibinfo {author} {\bibfnamefont {T.~W.}\
  \bibnamefont {Ebbesen}},\ }\href {\doibase 10.1103/PhysRevLett.109.013901}
  {\bibfield  {journal} {\bibinfo  {journal} {Phys. Rev. Lett.}\ }\textbf
  {\bibinfo {volume} {109}},\ \bibinfo {pages} {013901} (\bibinfo {year}
  {2012})}\BibitemShut {NoStop}%
\bibitem [{Note1()}]{Note1}%
  \BibitemOpen
  \bibinfo {note} {Postselection in optics is used in different contexts: It
  relates to correlated modes and conditioning (e.g. heralded single photon
  sources~\cite {Kok2007}), and to correlated properties in one mode (e.g.
  strong interference of orthogonally polarized fields by polarization
  projection~\cite {Hosten2008,Dennis2012,Gorodetski2012}). The usage here
  refers to the latter.}\BibitemShut {Stop}%
\bibitem [{\citenamefont {Jackson}(1999)}]{Jackson1999}%
  \BibitemOpen
  \bibfield  {author} {\bibinfo {author} {\bibfnamefont {J.~D.}\ \bibnamefont
  {Jackson}},\ }\href {\doibase 10.4006/1.3025509} {\emph {\bibinfo {title}
  {{Classical Electrodynamics}}}},\ \bibinfo {edition} {3rd}\ ed.\ (\bibinfo
  {publisher} {Wiley},\ \bibinfo {address} {New York},\ \bibinfo {year}
  {1999})\BibitemShut {NoStop}%
\bibitem [{\citenamefont {Dressel}(2015)}]{Dressel2015}%
  \BibitemOpen
  \bibfield  {author} {\bibinfo {author} {\bibfnamefont {J.}~\bibnamefont
  {Dressel}},\ }\href {\doibase 10.1103/PhysRevA.91.032116} {\bibfield
  {journal} {\bibinfo  {journal} {Phys. Rev. A}\ }\textbf {\bibinfo {volume}
  {91}},\ \bibinfo {pages} {032116} (\bibinfo {year} {2015})}\BibitemShut
  {NoStop}%
\bibitem [{\citenamefont {Lukosz}\ and\ \citenamefont
  {Kunz}(1977)}]{Lukosz1977b}%
  \BibitemOpen
  \bibfield  {author} {\bibinfo {author} {\bibfnamefont {W.}~\bibnamefont
  {Lukosz}}\ and\ \bibinfo {author} {\bibfnamefont {R.~E.}\ \bibnamefont
  {Kunz}},\ }\href@noop {} {\bibfield  {journal} {\bibinfo  {journal} {J. Opt.
  Soc. Am.}\ }\textbf {\bibinfo {volume} {67}},\ \bibinfo {pages} {1615}
  (\bibinfo {year} {1977})}\BibitemShut {NoStop}%
\bibitem [{\citenamefont {Lee}\ \emph {et~al.}(2007)\citenamefont {Lee},
  \citenamefont {Kihm}, \citenamefont {Kihm}, \citenamefont {Choi},
  \citenamefont {Kim}, \citenamefont {Ropers}, \citenamefont {Park},
  \citenamefont {Yoon}, \citenamefont {Choi}, \citenamefont {Woo},
  \citenamefont {Kim}, \citenamefont {Lee}, \citenamefont {Park}, \citenamefont
  {Lienau},\ and\ \citenamefont {Kim}}]{Lee2007}%
  \BibitemOpen
  \bibfield  {author} {\bibinfo {author} {\bibfnamefont {K.~G.}\ \bibnamefont
  {Lee}}, \bibinfo {author} {\bibfnamefont {H.~W.}\ \bibnamefont {Kihm}},
  \bibinfo {author} {\bibfnamefont {J.~E.}\ \bibnamefont {Kihm}}, \bibinfo
  {author} {\bibfnamefont {W.~J.}\ \bibnamefont {Choi}}, \bibinfo {author}
  {\bibfnamefont {H.}~\bibnamefont {Kim}}, \bibinfo {author} {\bibfnamefont
  {C.}~\bibnamefont {Ropers}}, \bibinfo {author} {\bibfnamefont {D.~J.}\
  \bibnamefont {Park}}, \bibinfo {author} {\bibfnamefont {Y.~C.}\ \bibnamefont
  {Yoon}}, \bibinfo {author} {\bibfnamefont {S.~B.}\ \bibnamefont {Choi}},
  \bibinfo {author} {\bibfnamefont {D.~H.}\ \bibnamefont {Woo}}, \bibinfo
  {author} {\bibfnamefont {J.}~\bibnamefont {Kim}}, \bibinfo {author}
  {\bibfnamefont {B.}~\bibnamefont {Lee}}, \bibinfo {author} {\bibfnamefont
  {Q.~H.}\ \bibnamefont {Park}}, \bibinfo {author} {\bibfnamefont
  {C.}~\bibnamefont {Lienau}}, \ and\ \bibinfo {author} {\bibfnamefont {D.~S.}\
  \bibnamefont {Kim}},\ }\href {\doibase 10.1038/nphoton.2006.37} {\bibfield
  {journal} {\bibinfo  {journal} {Nat. Photon.}\ }\textbf {\bibinfo {volume}
  {1}},\ \bibinfo {pages} {53} (\bibinfo {year} {2007})}\BibitemShut {NoStop}%
\bibitem [{\citenamefont {Neugebauer}\ \emph {et~al.}(2014)\citenamefont
  {Neugebauer}, \citenamefont {Bauer}, \citenamefont {Banzer},\ and\
  \citenamefont {Leuchs}}]{Neugebauer2014}%
  \BibitemOpen
  \bibfield  {author} {\bibinfo {author} {\bibfnamefont {M.}~\bibnamefont
  {Neugebauer}}, \bibinfo {author} {\bibfnamefont {T.}~\bibnamefont {Bauer}},
  \bibinfo {author} {\bibfnamefont {P.}~\bibnamefont {Banzer}}, \ and\ \bibinfo
  {author} {\bibfnamefont {G.}~\bibnamefont {Leuchs}},\ }\href
  {http://pubs.acs.org/doi/abs/10.1021/nl5003526} {\bibfield  {journal}
  {\bibinfo  {journal} {Nano Lett.}\ }\textbf {\bibinfo {volume} {14}},\
  \bibinfo {pages} {2546} (\bibinfo {year} {2014})}\BibitemShut {NoStop}%
\bibitem [{\citenamefont {Neugebauer}\ \emph {et~al.}(2015)\citenamefont
  {Neugebauer}, \citenamefont {Bauer}, \citenamefont {Aiello},\ and\
  \citenamefont {Banzer}}]{Neugebauer2015}%
  \BibitemOpen
  \bibfield  {author} {\bibinfo {author} {\bibfnamefont {M.}~\bibnamefont
  {Neugebauer}}, \bibinfo {author} {\bibfnamefont {T.}~\bibnamefont {Bauer}},
  \bibinfo {author} {\bibfnamefont {A.}~\bibnamefont {Aiello}}, \ and\ \bibinfo
  {author} {\bibfnamefont {P.}~\bibnamefont {Banzer}},\ }\href {\doibase
  10.1103/PhysRevLett.114.063901} {\bibfield  {journal} {\bibinfo  {journal}
  {Phys. Rev. Lett.}\ }\textbf {\bibinfo {volume} {114}},\ \bibinfo {pages}
  {063901} (\bibinfo {year} {2015})}\BibitemShut {NoStop}%
\bibitem [{\citenamefont {Stalder}\ and\ \citenamefont
  {Schadt}(1996)}]{Stalder1996a}%
  \BibitemOpen
  \bibfield  {author} {\bibinfo {author} {\bibfnamefont {M.}~\bibnamefont
  {Stalder}}\ and\ \bibinfo {author} {\bibfnamefont {M.}~\bibnamefont
  {Schadt}},\ }\href {\doibase 10.1364/OL.21.001948} {\bibfield  {journal}
  {\bibinfo  {journal} {Opt. Lett.}\ }\textbf {\bibinfo {volume} {21}},\
  \bibinfo {pages} {1948} (\bibinfo {year} {1996})}\BibitemShut {NoStop}%
\bibitem [{\citenamefont {Lieb}\ \emph {et~al.}(2004)\citenamefont {Lieb},
  \citenamefont {Zavislan},\ and\ \citenamefont {Novotny}}]{Lieb2004}%
  \BibitemOpen
  \bibfield  {author} {\bibinfo {author} {\bibfnamefont {M.~A.}\ \bibnamefont
  {Lieb}}, \bibinfo {author} {\bibfnamefont {J.~M.}\ \bibnamefont {Zavislan}},
  \ and\ \bibinfo {author} {\bibfnamefont {L.}~\bibnamefont {Novotny}},\ }\href
  {\doibase 10.1364/JOSAB.21.001210} {\bibfield  {journal} {\bibinfo  {journal}
  {J. Opt. Soc. Am. B}\ }\textbf {\bibinfo {volume} {21}},\ \bibinfo {pages}
  {1210} (\bibinfo {year} {2004})}\BibitemShut {NoStop}%
\bibitem [{\citenamefont {Foreman}\ \emph {et~al.}(2008)\citenamefont
  {Foreman}, \citenamefont {Romero},\ and\ \citenamefont
  {T{\"{o}}r{\"{o}}k}}]{Foreman2008}%
  \BibitemOpen
  \bibfield  {author} {\bibinfo {author} {\bibfnamefont {M.~R.}\ \bibnamefont
  {Foreman}}, \bibinfo {author} {\bibfnamefont {C.~M.}\ \bibnamefont {Romero}},
  \ and\ \bibinfo {author} {\bibfnamefont {P.}~\bibnamefont
  {T{\"{o}}r{\"{o}}k}},\ }\href
  {http://www.opticsinfobase.org/ol/fulltext.cfm?uri=ol-33-9-1020{\&}id=157550}
  {\bibfield  {journal} {\bibinfo  {journal} {Opt. Lett.}\ }\textbf {\bibinfo
  {volume} {33}},\ \bibinfo {pages} {1020} (\bibinfo {year}
  {2008})}\BibitemShut {NoStop}%
\bibitem [{\citenamefont {Kok}\ \emph {et~al.}(2007)\citenamefont {Kok},
  \citenamefont {Munro}, \citenamefont {Nemoto}, \citenamefont {Ralph},
  \citenamefont {Dowling},\ and\ \citenamefont {Milburn}}]{Kok2007}%
  \BibitemOpen
  \bibfield  {author} {\bibinfo {author} {\bibfnamefont {P.}~\bibnamefont
  {Kok}}, \bibinfo {author} {\bibfnamefont {W.~J.}\ \bibnamefont {Munro}},
  \bibinfo {author} {\bibfnamefont {K.}~\bibnamefont {Nemoto}}, \bibinfo
  {author} {\bibfnamefont {T.~C.}\ \bibnamefont {Ralph}}, \bibinfo {author}
  {\bibfnamefont {J.~P.}\ \bibnamefont {Dowling}}, \ and\ \bibinfo {author}
  {\bibfnamefont {G.~J.}\ \bibnamefont {Milburn}},\ }\href {\doibase
  10.1103/RevModPhys.79.135} {\bibfield  {journal} {\bibinfo  {journal} {Rev.
  Mod. Phys}\ }\textbf {\bibinfo {volume} {79}},\ \bibinfo {pages} {135}
  (\bibinfo {year} {2007})}\BibitemShut {NoStop}%
\end{thebibliography}%
\end{document}